# EXCITATION FUNCTION OF THE $^{nat}$Ta(p,x)$^{178m2}$Hf REACTION

Yu. E. Titarenko, K. V. Pavlov, V. I. Rogov, A. Yu. Titarenko, S. N. Yuldashev, V. M. Zhivun, ITEP, Moscow, 117218 Russia;
A. V. Ignatyuk, IPPE, Obninsk, 249033 Russia;
S. G. Mashnik, LANL, Los Alamos, 87545 NM, USA;
S. Leray, A. Boudard, J.-C. David, D. Mancusi, CEA, Saclay, France;
J. Cugnon, University of Liege, Belgium;
Y. Yariv, Soreq NRC, Yavne, Israel;
K. Nishihara, N. Matsuda, JAEA, Tokai, Japan;
H. Kumawat, BARC, Mumbai, India & GSI, Darmstadt, Germany;
A. Yu. Stankovskiy, SCK•CEN, Belgium

*Abstract*

$^{178m2}$Hf is an extremely interesting isomeric state due to its potential energy capacity level. One possible way to obtain it is by irradiation of a $^{nat}$Ta sample with a high-current proton accelerator. Up to now, there was no information in the international experimental nuclear data base (EXFOR) for this reaction. Irradiations of $^{nat}$Ta samples performed for other purposes provide an opportunity to address this question. This paper presents the $^{172m2}$Hf independent production cross-sections determined by gamma-ray spectrometry. The $^{nat}$Ta(p,x)$^{172m2}$Hf excitation function is studied in the 20-3500 MeV energy range. Comparisons with results by several nuclear models (ISABEL, Bertini, INCL4.5+ABLA07, PHITS, CASCADE07, and CEM03.02) used as event-generators in modern transport codes are also reported. However, since such models are generally not able to separately predict ground and isomeric states of reaction products, only $^{178}$Hf independent and cumulative cross-section data are compared.

## SAMPLES IRRADIATION

Irradiation of 20 thin sample-targets of $^{nat}$Ta with protons in the energy range 40-2600 MeV have been performed at the ITEP accelerator during the period from September 1, 2006 to August 31, 2009, within the framework of the ISTC Project #3266. Figs. 1 and 2 show the external proton channel with a stand for irradiation of samples and a system for their transport from the magnetic hall to a laboratory room, respectively. Irradiation parameters of the samples are listed in Tab. 1.

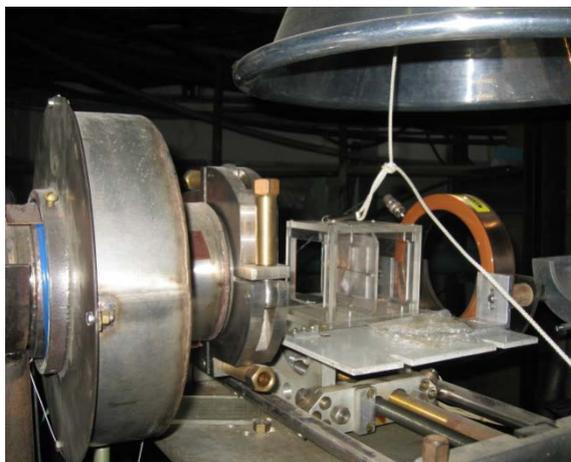

Fig. 1. External proton channel with a stand for irradiation of samples.

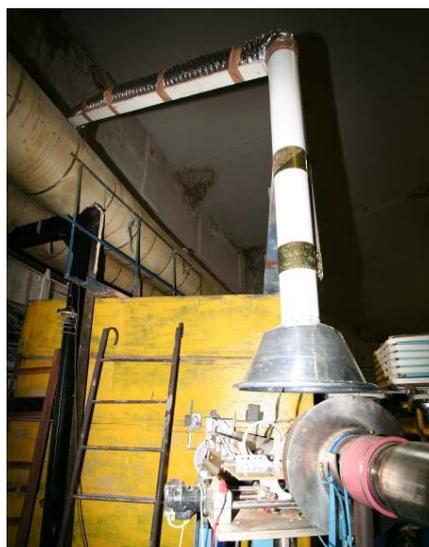

Fig. 2. Transport system for the irradiated samples.

Table 1. Irradiation parameters of the Ta samples.

| Proton energy (MeV) | Sample mass (mg) | Irradiation time (min) | Average proton flux $p/(cm^2 \cdot s) \times 10^{-10}$ | Proton energy (MeV) | Sample mass (mg) | Irradiation time (min) | Average proton flux $p/(cm^2 \cdot s) \times 10^{-10}$ |
|---|---|---|---|---|---|---|---|
| 2605±8 | 359.8 | 24.18 | 7.44 ± 0.64 | 399±2 | 353.0 | 1262 | 1.89±0.16 |
| 1598±4 | 354.2 | 29.12 | 5.74 ± 0.48 | 248±1 | 354.4 | 28.0 | 8.09 ± 0.58 |
| 1199±3 | 359.6 | 28.0 | 6.98 ± 0.56 | 148±1 | 355.0 | 27.0 | 5.17 ± 0.40 |
| | 355.0 | 2502 | 5.12±0.37 | | 127.7 | 31.0 | 6.84 ± 0.50 |
| 799±2 | 357.5 | 24.0 | 6.05 ± 0.49 | 97±1 | 358.6 | 38.5 | 3.23 ± 0.27 |
| | 358.4 | 3022 | 8.55±0.64 | | 123.8 | 30.0 | 5.13 ± 0.38 |
| 599±2 | 357.5 | 28.0 | 4.55 ± 0.54 | 66±1 | 354.8 | 65.5 | 1.81 ± 0.13 |
| | 355.5 | 759 | 7.97±0.57 | | 127.0 | 31.0 | 6.14 ± 0.43 |
| 399±2 | 355.8 | 23.0 | 4.21 ± 0.38 | 43±1 | 357.8 | 25.0 | 4.57 ± 0.34 |
| | | | | | 352.5 | 45.0 | 3.67 ± 0.26 |

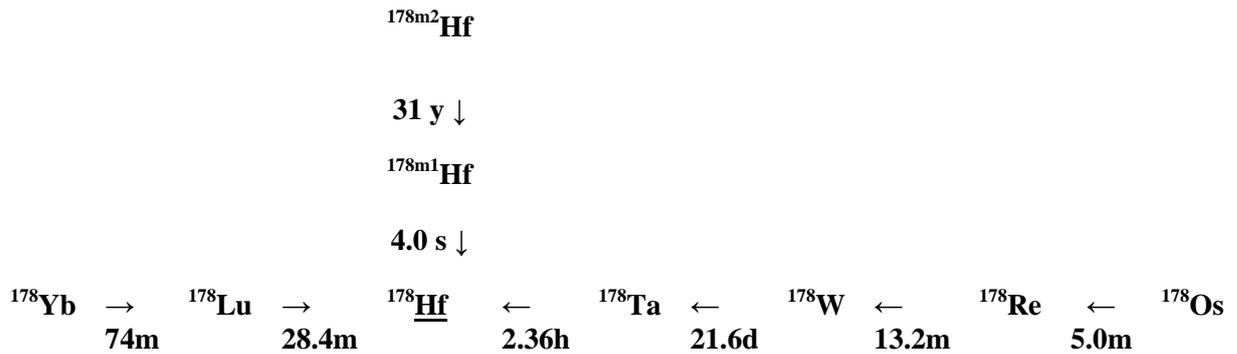

$^{178m2}$Hf

31 y ↓

$^{178m1}$Hf

4.0 s ↓

$^{178}$Yb → $^{178}$Lu → **$^{178}$Hf** ← $^{178}$Ta ← $^{178}$W ← $^{178}$Re ← $^{178}$Os
74m    28.4m    2.36h    21.6d    13.2m    5.0m

Fig. 3. Chain of decays for the production of $^{178}$Hf.

## GAMMA SPECTRA MEASUREMENT

During the period of this study, about 300 gamma- and alpha-spectra were measured, that allowed us to determine cross section for 882 residual product nuclei from our proton-induced reactions, which were presented as 173 excitation functions in Refs. [1-3].

However, some products of reactions that are of great interest for different applications have not been measured; among them are the beta emitters ($^3$H, $^{14}$C, $^{36}$Cl, ...) and the gaseous ($^{3,4}$He, ...). $^{178m2}$Hf, discussed in many papers (see [4] and references therein) as a potential "energy storage" state, is also not measured from this group.

As can be seen from Fig. 3, $^{178m2}$Hf is produced only in nuclear reactions, therefore its i(m2) cross section is much smaller than the total independent, i(m2+m1+g), cross section for the production of $^{178}$Hf, and especially smaller than the cumulative cross section (c).

In order to determine the excitation function for the production of $^{178m2}$Hf, the measurement of gamma spectra of early irradiated samples of $^{nat}$Ta was continued during 2012. A longer time after irradiation provided a significant decrease in the radioactivity background produced by the natural decay of the reaction products with shorter half-lives, did increase the detection limit, and allowed us to confidently identify the $^{178m2}$Hf by its gamma lines: 213.4 keV (81.4%), 216.7 keV (64.6%), and 574.2 keV (88.0%) [5]

Other gamma lines were not used in the current work, either because of their low quantum yield (< 20%) or because of the coincidence of their energies, within the resolution limit of the spectrometer, with the energies of $^{172}$Lu ($t_{1/2}$ = 6.70 d) gamma lines, which is in equilibrium with the long-lived $^{172}$Hf ($t_{1/2}$ = 1.87 y), as well as because of their location in the x-ray range (<100 keV). Coincidences with background gamma lines in these ranges were not found.

The measured gamma spectrum of the $^{nat}$Ta sample, in about 5 years after irradiation with 600 MeV protons is shown in Fig. 4. The background spectrum of the laboratory room in which the measurement was made is shown in Fig. 5. The methodology and formulas for determining the cross sections of nuclear reaction products using the gamma spectrometry are described in detail in Refs. [1-3].

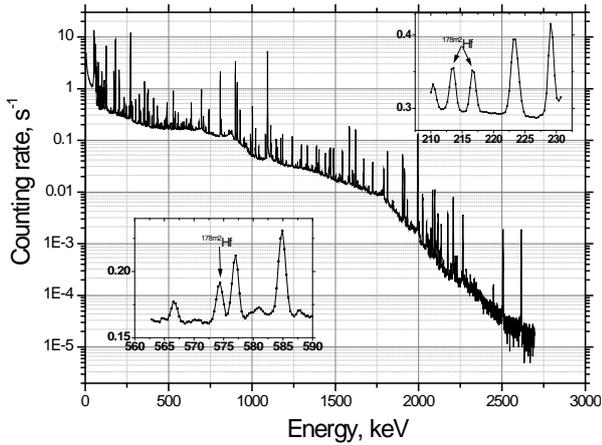

Fig. 4. The measured gamma spectrum of the natTa sample, in about 5 years after irradiation with 600 MeV protons. The measuring time was 7 days.

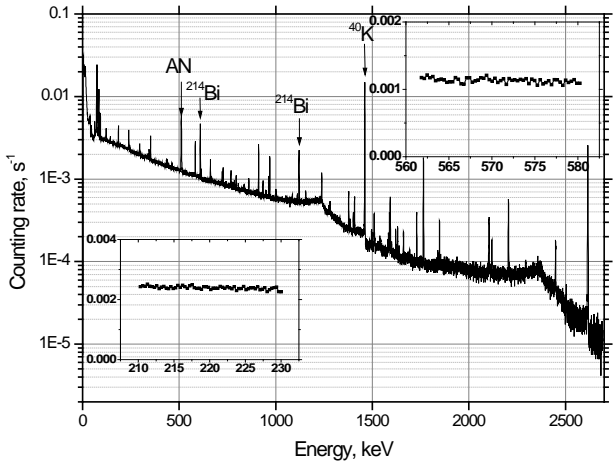

Fig. 5. The laboratory room background spectrum. The measuring duration was 14 days

## EXPERIMENTAL RESULTS

The measured values of the independent cross sections for $^{178m2}$Hf ($\sigma_{i(m2)}^{178m2Hf}$) are shown using a linear scale in Fig. 6 together with similar data measured earlier at JINR, Dubna by Karamian et al. [6]. Despite the fact that in Ref. [6] thick targets were irradiated (a = 2 cm), their result at a proton energy of 660 MeV agrees within the experimental error with our value, since the energy loss is of only ~ 7%. Discrepancies for energies of 200 MeV and 100 MeV are due to bigger losses of energy (up to the beam stop).

## COMPARISON WITH CALCULATIONS

Theoretical independent ($\sigma_{i(m1+m2+g)}^{178Hf}$) and cumulative ($\sigma_c^{178Hf}$) cross sections (yields) calculated with nuclear reaction models ISABEL, Bertini, INCL4.5+ABLA07, PHITS, CASCADE07, and CEM03.02, used as event-generators in modern high-energy transport codes, are shown with a logarithmic scale in Fig. 7 together with the experimental values for the independent cross section (yield) $^{178m2}$Hf ($\sigma_{i(m2)}^{178m2Hf}$).

All the listed above nuclear reaction models, described in some detail in Ref. [3], can simulate only independent cross sections (yields) of the residual nuclei products, but do not calculate the yields of their possible isomeric states.

## DISCUSSION AND CONCLUSION

In neutron-induced reactions, the yield of isomer $^{178m2}$Hf ($J^\pi=16^+$) was analyzed in Refs. [7-9]. For the neutron energy of 14.8 MeV, the estimated yield of $^{178m2}$Hf in the reaction $^{179}$Hf (n,2n) corresponds to the value of 6.33 ± 0.28 mb [8]. Since the total cross section for the $^{179}$Hf(n,2n) reaction at this energy is equal to 2200 ± 150 mb, the value of the isomer ratio is equal ~ 0.003 with an uncertainty of ~ 10%. On the other hand, the empirical systematics of isomeric ratios for neutron-induced reactions provides a value of 0.001 with an accuracy of more than 50%, for the 16$^+$ spin case [7].

Theoretical calculations [9] of the isomeric ratios for the high-spin states in the neutron energy up to 20 MeV yield values, on average, consistent with the systematics [7]. Note that the calculations indicate a systematic increase in the isomeric ratios at energies above 15 MeV, but such changes depend significantly on the scheme of the low-lying levels of the nucleus and on angular momenta excited in reactions. If to adopt as an estimate of the isomer ratio the value of 0.001-0.003, then for the total cross sections shown in Fig. 7 at energies below 100 MeV, we get for the $^{178m2}$Hf isomer yield a value of 0.001-0.06 mb.

This is approximately about ten or more times lower than the values obtained in our experiment. This discrepancy can be attributed to the energy dependence of the isomer ratio. Unfortunately, a significant absence of low-lying states in the scheme of $^{178}$Hf severely limits the reliability of the calculations of the isomeric ratio for the high-spin isomer $16^+$. Therefore, the accumulation of experimental data for the production cross section of this isomer in various reactions is of considerable interest both for potential applications as well as for further developments of theoretical models of settlement of high-spin states.

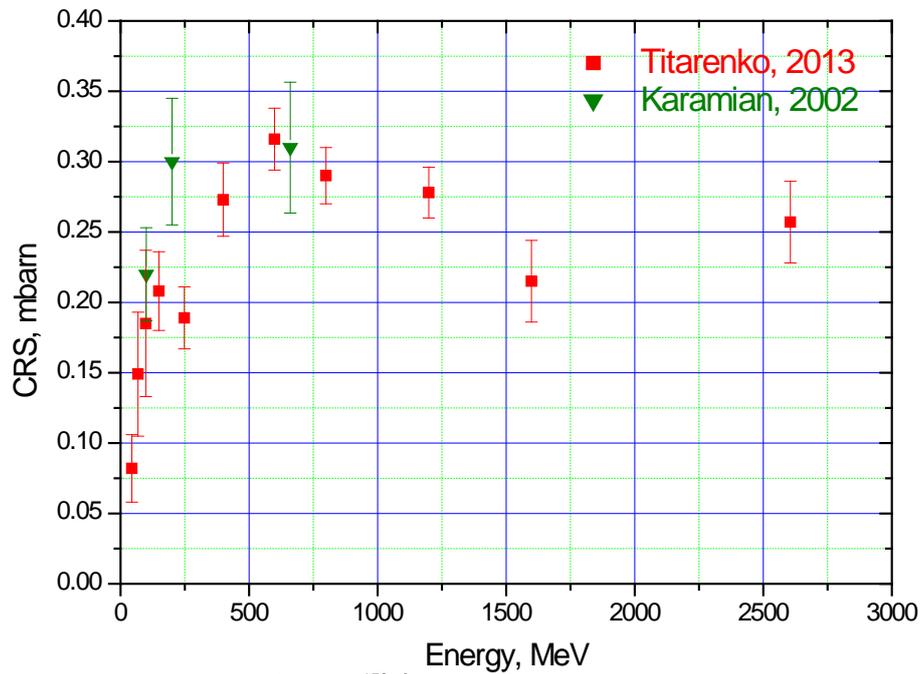

Fig. 6. Experimental excitation function $^{nat}Ta(p,x)^{178m2}Hf_{i\,(m2)}$ measured in the current work at ITEP (■) and previous values from JINR (▼) [6].

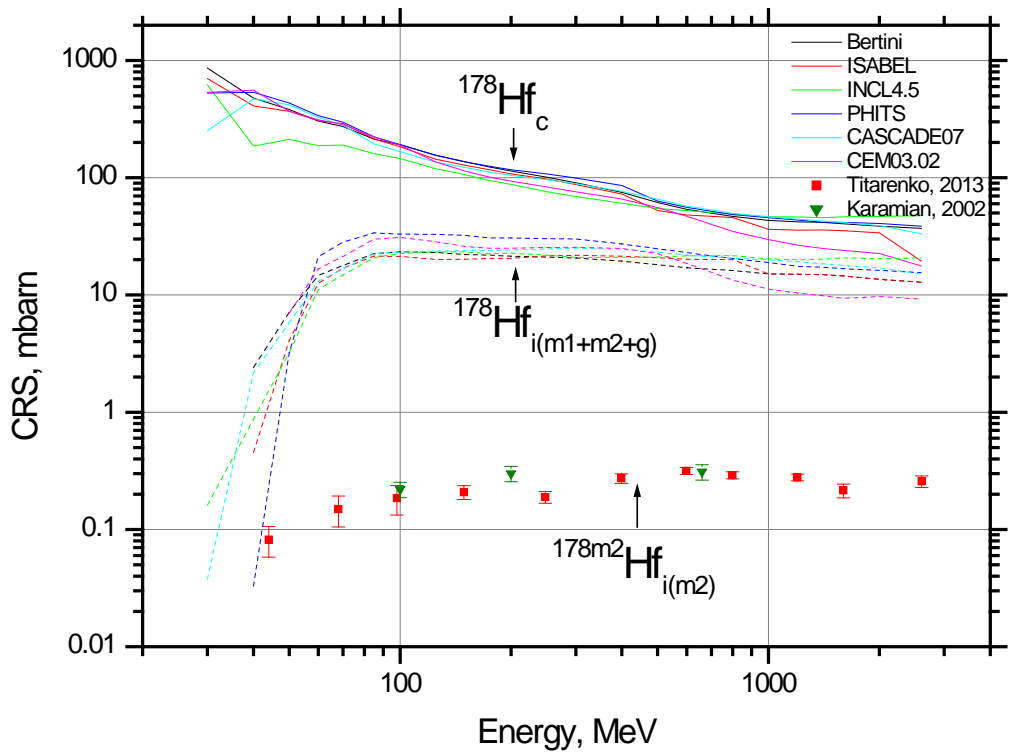

Fig. 7. Theoretical excitation functions for the cumulative (c) and individual (i) production of $^{178}Hf$, $^{178}Hf_c$ and $^{178}Hf_{i(m2+m1+g)}$, respectively, compared with the experimental excitation function for the production of $^{178m2}Hf_{i\,(m2)}$.

## ACKNOWLEDGEMENT

We thank Dr. Roger Martz for a careful reading of our manuscript and several useful suggestions. This work was supported by the International Science and Technology Center, project No. 3266, by the State Corporation Rosatom and National Research Centre Kurchatov Institute. Part of the work performed at LANL was carried out under the auspices of the National Nuclear Security Administration of the U.S. Department of Energy at Los Alamos National Laboratory under Contract No. DE-AC52-06NA25396.